\documentclass{article}
\usepackage[english]{babel}
\usepackage[utf8]{inputenc}
\usepackage[T1]{fontenc} 
\usepackage{authblk}
\usepackage[a4paper, margin=1.2in]{geometry}
\usepackage{epsfig,endnotes}
\usepackage{epigraph}
\usepackage{subcaption}
\usepackage{todonotes} 
\usepackage[hyphens]{url}
\usepackage[bb=pazo]{mathalpha}
\usepackage{pgfplots}
\usepackage[table]{xcolor}
\usepackage{booktabs}
\pgfplotsset{compat=1.18}
\usepackage{pgfplotstable}

\definecolor{color1}{HTML}{003f5c}
\definecolor{color2}{HTML}{58508d}
\definecolor{color3}{HTML}{bc5090}

\begin{document}

\title{Just in Plain Sight: Unveiling CSAM Distribution Campaigns on the Clear Web}

\author[1]{Nikolaos Lykousas}
\author[2,3]{Constantinos Patsakis}

\affil[1]{Data Centric, Romania}
\affil[2]{Department of Informatics, University of Piraeus, 80 Karaoli \& Dimitriou str., 18534 Piraeus, Greece}
\affil[3]{Information Management Systems Institute of Athena Research Centre, Greece}

\date{}
\maketitle


\maketitle
\begin{abstract}
    Child sexual abuse is among the most hideous crimes, yet, after the COVID-19 pandemic, there is a huge surge in the distribution of child sexual abuse material (CSAM). Traditionally, the exchange of such material is performed on the dark web, as it provides many privacy guarantees that facilitate illicit trades. However, the introduction of end-to-end encryption platforms has brought it to the deep web. In this work, we report our findings for a campaign of spreading child sexual abuse material on the clear web. The campaign utilized at least 1,026 web pages for at least 738,286 registered users. Our analysis details the operation of such a campaign, showcasing how social networks are abused and the role of bots, but also the bypasses that are used. Going a step further and exploiting operational faults in the campaign, we gain insight into the demand for such content, as well as the dynamics of the user network that supports it.     
\end{abstract}

\section{Introduction}

Traditionally, communities of perpetrators producing, consuming, and trading child sexual abuse material (CSAM) have been studied primarily within the context of the dark web. The veil of anonymity, along with the lack of censorship and inherent confidentiality of dark web services, facilitates, among others, such illicit exchanges \cite{guitton2013review,gannon2023child,ngo2024discovering}. During the COVID-19 pandemic, CSAM distribution has increased significantly \cite{iocta}; however, this period coincided with the widespread adoption of messengers supporting end-to-end encryption. The latter is a great enabler for distributing such content, as the material can be shared with a far wider audience without the need for special software. Moreover, it facilitates the exchange and consumption of such content from mobile devices whose use has become ubiquitous.

It should be emphasized that many such communities follow pyramid schemes that follow a hierarchy based on content ownership. Thus, many members are requested to provide new content to be allowed. On the other hand, many of the members simultaneously seek to engage and recruit more users by spamming networks with CSAM. This behavior can be attributed to their need to normalize the consumption of such content, but also to attract more people to the group, who will generate more CSAM that they will consume. As a result, there are often reports about CSAM being distributed via traditional social networks and, of course, users grooming minors over various online platforms.

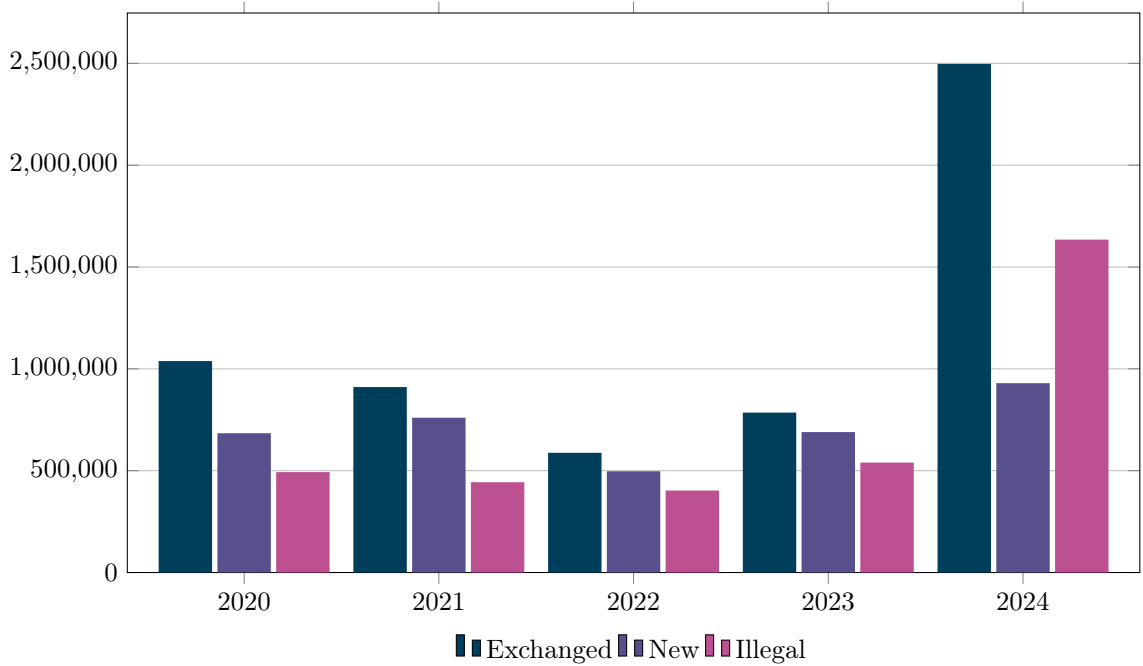
\begin{figure}[th]
\begin{tikzpicture}
\begin{axis}[
    ybar,
    ymajorgrids=true,
    bar width=20pt,
    width=\columnwidth,
    height=9cm,
    symbolic x coords={2020, 2021, 2022, 2023, 2024},
    xtick=data,
    legend style={at={(0.5,-0.1)},
    anchor=north, legend columns=3,draw=none},
    enlarge x limits=0.15,
    ymin=0,
    scaled y ticks=false,
    yticklabel style={
        /pgf/number format/fixed,
        /pgf/number format/precision=5
    },
]

\addplot[fill=color1, draw=none] coordinates {
    (2020,1038250)
    (2021,910642)
    (2022,587852)
    (2023,785322)
    (2024,2497438)
};

\addplot[fill=color2, draw=none] coordinates {
    (2020,683890)
    (2021,760054)
    (2022,497001)
    (2023,689523)
    (2024,929733)
};

\addplot[fill=color3, draw=none] coordinates {
    (2020,492961)
    (2021,443391)
    (2022,402601)
    (2023,539836)
    (2024,1634636)
};

\legend{Exchanged, New, Illegal}
\end{axis}
\end{tikzpicture}
\caption{Reported CSAM record growth according to INHOPE \cite{inhope}.}
\label{fig:inhope}
\end{figure}

Unfortunately, the problem of CSAM distribution has reached unprecedented levels. According to the latest report from INHOPE \cite{inhope}, not only has the reported suspicious content skyrocketed, but the illegal content has actually tripled that of the previous year, as shown in Figure \ref{fig:inhope}. Moreover, as can be observed in the figure, there is a steady influx of new content, meaning that more children are subjected to this torture. Most of the content reported to INHOPE is shared in forums (61.07\,\%), followed by image hosting services (20.86\,\%), websites 14.86\,\%, and file hosting services (2.03\,\%).

The numbers of the recent takedown of Kidflix are also shocking \cite{kidflix}. Around 1.8 million users worldwide had logged onto the platform between April 2022 and March 2025. When seized, the server contained approximately 72,000 videos, while the platform had around 91,000 videos with an estimated total duration of 6,288 hours.

Contrary to most studies in the literature related to CSAM distribution, which study the availability of such content, the distribution methods, as well as the geographic distribution of the users \cite{wortley2024accessing}, we study a wide online campaign on the clear web that used 1,026 unique domains. Our study goes beyond statistics by providing insight into the modus operandi of this campaign, detailing how and where the content is distributed. Moreover, we shed light on the dynamics of the referral network, reporting the inviter/invitee dynamics and some characteristics of the CSAM users, who worldwide are on the scale of 738,286. Despite its content and the fact that it is operated on the clear web, it has been ongoing since 2022. It abuses various platforms and is orchestrated through Telegram and other clear web channels to communicate with potential users and buyers, but also to share part of the material. In compliance with ethical standards and legal frameworks, we highlight that we did not access, view, or download any CSAM. Our analysis was limited to metadata, referral mechanisms, domain activity, and platform behavior. While this ensures safety and legality, it limits our ability to verify the exact nature of the shared content.

To this end, the following section presents an overview of the related work. Then, in Section \ref{sec:overview}, we provide an overview of the CSAM campaign, and in Section \ref{sec:methodology}, we document the data collection methodology employed for discovering and extracting information from URLs hosting instances of CSAM distribution platforms on the clear web, while also shedding light on the common mechanics used by their operators. We then examine the methods by which these campaigns are disseminated on $\mathbb{X}$/Twitter, focusing on the use of sockpuppet accounts and compromised user profiles. Furthermore, in Section \ref{sec:network}, we analyze referral network data extracted from these platforms to assess the extent to which the MLM structure is organic, consisting of actual CSAM users who invite other users, perpetuating the cycle.
Subsequently, in Section \ref{sec:consumers} we leverage anonymized data captured during a brief window of opportunity, when one of the platforms inadvertently leaked information about visitors' IP addresses and web browsers. Using this data, we provide insights into the visitors' browser footprints, geographic distribution, and other relevant patterns. Since our research has its limitations, we discuss them in Section \ref{sec:limitations}. Finally, the article concludes, giving an overview of our contributions and analysis, and discussing ideas for future work and mitigation measures.

\section{Related work} 

Due to the fact that exchanging CSAM is illegal, individuals engaging in such exchanges attempt to implement as many security measures as possible. To this end, users tend to use networks that provide them with as much privacy as possible. 

The most obvious method to hide this traffic is through the use of the dark web. Essentially, the traffic in Tor and I2P, the primary dark web networks, not only encrypt all the exchanged traffic but also hide the traffic by, e.g., forwarding the traffic from one node to another, and then using exit nodes to obfuscate where the traffic originates. As a result, dark web services are repeatedly reported for hosting and serving CSAM \cite{liggett2020dark,woodhams2021characteristics,van2022even,kidflix}, while some search engines are also reported to promote it \cite{nurmi2024investigating}.

Moreover, peer-to-peer networks offer a veil of anonymity, and the fact that files are broken into smaller chunks that are exchanged among peers facilitates the obfuscation of these exchanges, but also some sort of deniability. Therefore, BitTorrent has been widely used to exchange CSAM \cite{rutgaizer2012detecting,shavitt2013child,wolak2014measuring} with several methods and tools introduced to detect such content, e.g., file names \cite{panchenko2012detection}. Nevertheless, other P2P networks, such as IPFS, have been reported to host CSAM \cite{sokoto2024guardians}.

Finally, end-to-end encryption communication platforms are increasingly
being abused for CSAM exchange and communication \cite{iocta2}. Beyond strong cryptographic primitives, these platforms provide secrecy and several privacy guarantees in a very compact form, as they can operate in mobile devices, without the need for any expertise.

Given that the detection of most CSAM is based on hashes, the detection of new content is very challenging, e.g., the hash can easily be manipulated to significantly deviate from the original by simply cropping, rotating, or blurring an image or a video frame. Therefore, further research in this direction is necessary, considering the role of AI and machine learning in this area and the potential to automate the corresponding pipelines \cite{singh2024role}. For instance, similar to \cite{panchenko2012detection}, Guerra
and Westlake \cite{guerra2021detecting} use the names on the websites to detect CSAM, or age estimation \cite{castrillon2018evaluation} to quickly assess the content. Nevertheless, the elephant in the room remains: the identification of victims and offenders.

According to \cite{insoll2022risk}, almost half of the sample of CSAM users responded that after watching CSAM, they then sought direct contact with children through online platforms, signifying the impact that watching such material has on its users. In particular, most users reported that they had accidentally been exposed to this material for the first time, usually at a young age. The latter is also aligned with previous studies \cite{wortley2024accessing,nurmi2024investigating}.

\section{Overview of the CSAM campaign}
\label{sec:overview}
The CSAM campaign that we examine follows a Multi-level Marketing (MLM) scheme. In essence, the operators employ a `freemium' model, offering some material for free while reserving more explicit content for higher tiers.
The tiers are unlocked either through payments or by inviting other users through personalized referral links. The tiers unlocked with a price that ranges from \$20 to \$120, and according to the platform, provide access to even several terabytes of content.

In this regard, the following paragraphs outline the two main pillars of this campaign, namely the hosting and the distribution of the invitations.

As discussed, the campaign utilizes the clear web, employing multiple domains to host web platforms for user registration and content access. The domains typically have a particularly short lifespan, likely intended to evade attempts by law enforcement agencies to systematically eliminate, investigate, and prosecute perpetrators. Moreover, they use a specific template that gradually changes. 
A unique characteristic of the campaign is the choice of the domain names. The operators use domain names that exhibit characteristics of Domain Generation Algorithms (DGA) \cite{plohmann2016comprehensive,casinoLHPH21}. Practically, the domain name in most cases is a combination of random letters and a few digits, with a preference for specific TLDs, as discussed later.

Furthermore, the illicit material distributed through these platforms is typically not hosted on the same servers as the platforms themselves but instead stored on various cloud upload services and content delivery networks (CDNs). 

The referral system of the campaign resembles traditional social media referral mechanisms, enabling the rapid and systematic dissemination across mainstream platforms, such as $\mathbb{X}$, Facebook, and Instagram. Moreover, black-hat marketing methods, such as sockpuppets or posts from compromised social media accounts, further amplify the spread to a broader audience. Interestingly, our analysis demonstrates that while the spread of the campaign is orchestrated by its operators to reach potential buyers, a significant portion of the traffic to these platforms is organic. This traffic originates from offenders actively seeking to unlock higher content tiers by sharing their personalized invite links. To this end, sites that allow for posting without any form of authorization or CAPTCHA mechanism are frequently abused. These include paste-like platforms\footnote{e.g. \url{https://archive.ph/8Annc}} or comment sections on websites where the posted content is indexed by Google and becomes discoverable by offenders searching for specific keywords. 
An example of such a site is \url{updownradar.com}. Such keywords often include the initials \textit{``C''} and \textit{``P''}, a coded reference to child pornography, disguised as terms like \textit{`Club Penguin'} (a discontinued massively multiplayer online game) or phrases such as \textit{`Caldo de Pollo'} (Spanish for chicken soup). These keywords serve as covert language to avoid detection while used in plain sight. This technique has also been observed in the context of sexually grooming minors in Social Live Streaming Services that employ text-based moderation mechanisms \cite{LykousasP21}. Notably, the operators of this campaign actively prompt users to amplify and exacerbate its spread by promoting it in popular social platforms, through posting their referral links and using hashtags including such coded terms (i.e., \textit{\#caldodepollo}).  

Another aspect of interest for the specific campaign is the use of AI-generated media and images with embedded text for advertising the specific addresses where the campaign's platforms are hosted, as well as the Telegram handles of the campaign operators. The fact that these images are AI-generated is demonstrated by the fact that they have obvious faults, e.g., more fingers. These images are used both to promote the campaign on social platforms such as $\mathbb{X}$ (possibly to avoid posting links directly and thus evading detection/moderation mechanisms), and within the CSAM sites themselves, as logos or banners.

It must be noted that the first documented observation of this campaign that we have found can be traced back to a Reddit thread\footnote{\url{https://archive.ph/m2b8J}} in December 2022, where a user provided some initial insights\footnote{\url{https://archive.ph/G7IXa}} into the modus operandi of the clear-web CSAM distribution platforms:
\begin{quote}
    \textit{``They spam themselves on Twitter and dead subreddits and it's structured like a pyramid scheme (kinda). Basically they offer "rewards" in a tiering system if you invite people to their discord, from there it's filled with links to other discords, their telegram and crypto addresses. Ultimately the invites probably don't matter since the "free" users will likely never get much but the whales who contact them via telegram and actually start paying are the real prize. They can get taken down on Twitter all they like but the paying customers get to private channels and the pedos get their money. \\
I'm assuming they've avoided companies cracking down on them by sheer volume and the fact they tend to use various Russian characters or vague names to disguise their activities.''}
\end{quote}

\section{Data collection and processing methodology}
\label{sec:methodology}
As the spread of campaign links is highly dynamic, with perpetrators leveraging multiple outlets to post and promote their links, we adopt a two-faceted approach to collect URLs hosting such platforms.
To capture links posted on paste-like sites or other websites, as previously noted, we constructed a list of queries tailored to match some of the most evidently prominent TLDs by operators to host such platforms. These queries included relevant URL patterns for registering with referral links, such as \texttt{``.\{cc,de,ru,pw,ws,xyz,city\}/invite/i=''}. Using SerpAPI\footnote{\url{https://serpapi.com}}, we extracted results indexed by Google that matched these patterns.

Additionally, to retrieve relevant posts from $\mathbb{X}$ (formerly Twitter) that include hashtags selected by campaign operators for disseminating links (e.g., \texttt{\#caldodepollo}, \texttt{\#clubpenguin}, \texttt{\#irlixli}, \texttt{\#t33n}), we scraped data from the \texttt{sotwe}\footnote{\url{https://www.sotwe.com}} platform. This platform utilizes the Twitter API, alleviating the need for costly API access required to conduct such research \cite{ledford2023researchers}.

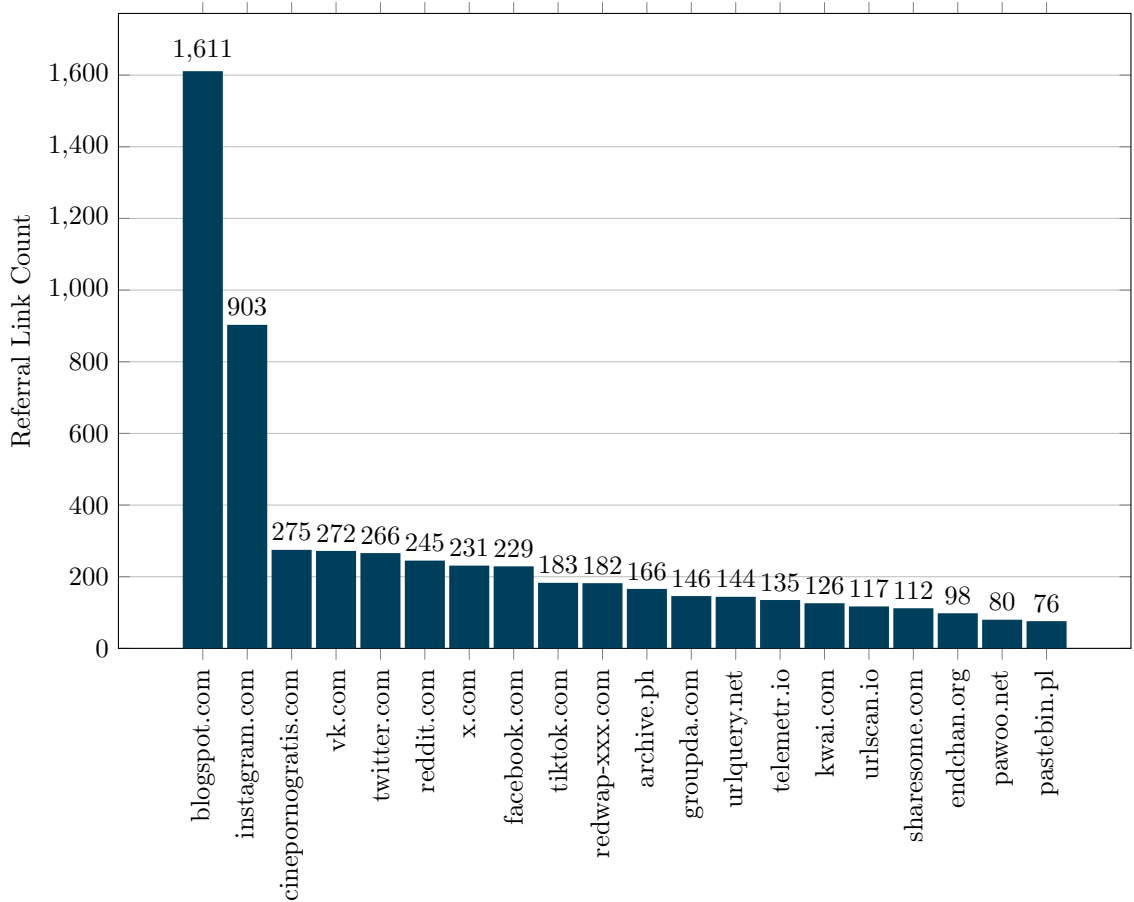
\begin{figure*}[!htb]
    \centering
    \begin{tikzpicture}
\begin{axis}[
    ybar,
    ymajorgrids=true,
    bar width=15pt,
    width=\textwidth,
    height=10cm,
    ylabel={Referral Link Count},
    symbolic x coords={
        blogspot.com, instagram.com, cinepornogratis.com, vk.com, twitter.com, reddit.com, x.com, facebook.com, tiktok.com, redwap-xxx.com,
        archive.ph, groupda.com, urlquery.net, telemetr.io, kwai.com, urlscan.io, sharesome.com, endchan.org, pawoo.net, pastebin.pl
    },
    xtick=data,
    xticklabel style={rotate=90, anchor=east},
    nodes near coords,
    ymin=0,
]

\addplot [fill=color1, draw=none] coordinates {
    (blogspot.com,1611)
    (instagram.com,903)
    (cinepornogratis.com,275)
    (vk.com,272)
    (twitter.com,266)
    (reddit.com,245)
    (x.com,231)
    (facebook.com,229)
    (tiktok.com,183)
    (redwap-xxx.com,182)
    (archive.ph,166)
    (groupda.com,146)
    (urlquery.net,144)
    (telemetr.io,135)
    (kwai.com,126)
    (urlscan.io,117)
    (sharesome.com,112)
    (endchan.org,98)
    (pawoo.net,80)
    (pastebin.pl,76)
};

\end{axis}
\end{tikzpicture}
    \caption{Top domains hosting CSAM referral links from Google results via SerpAPI.}
    \label{fig:serpapi_results}
\end{figure*}

Using the methods described above, we constructed a data collection pipeline to identify and extract referral links from these two sources, and subsequently collect the entire source code of the CSAM distribution platforms (if working at crawl time). The automation not only facilitates data processing, but it also prevents access to CSAM content. This pipeline was executed weekly over nine months (June 2024–March 2025), during which we identified 1,026 unique domains hosting various variants of CSAM distribution platforms. Specifically, the SerpAPI approach yielded 745 unique domains, while the sotwe-based approach yielded 281 more domains. It should be noted that for the second approach, almost all domains were not provided in text but embedded in images. To extract them, we employed an OCR pipeline similar to the one described in \cite{casino2023analysis}.  
We observe that while some domain names seem algorithmically generated, others appear to be hand-picked by the campaign operators, since they include words like \texttt{young}, \texttt{teen}, etc. Figure~\ref{fig:serpapi_results} shows the most popular sites indexed by Google containing referral links to CSAM distribution sites. We observe that sites allowing users to post comments are frequently abused, including the popular social media platforms, as well as pornographic and pasting websites. The latter finding is a common pattern of disseminating links and information in cybercrime contexts \cite{chertoff2015impact,madarie2019stolen}. Notably, the presence of sites such as \texttt{urlquery.net} and \texttt{urlscan.io} suggests that this campaign has already been flagged and investigated by other researchers or cybersecurity professionals.

\section{Campaign spread on $\mathbb{X}$ insights}
\label{sec:spread}
In total, our sotwe approach yielded 225,248 tweets containing one of the keywords that are known to be used by the campaign operators, which were posted by 1,616 unique users. In order to assess the extent to which such tweets were posted in an automated manner, we establish the following threshold: if the same account posted at least two tweets in our datasets within a window of 60 seconds, we label that account as a bot. Although this is a conservative threshold that likely underestimates the real scale of automation used to spread this campaign, it provides high confidence in our classification effort.

Using this criterion, we identified 731 automated accounts (45.2\,\% of all accounts) that were responsible for 223,980 tweets (99.4\,\% of the entire dataset). This means that for the duration of our data collection, the CSAM campaign was almost entirely spread by bots on the $\mathbb{X}$ platform. Notably, such bots posted an average of 306.4 tweets each, while non-automated accounts averaged just 2.0 tweets. Specifically, this observation suggests that the targeted hashtags were primarily used by CSAM campaign operators to communicate CSAM-hosting domains and Telegram channels used for monetization to potential users.

Examining the age distribution of automated accounts, we observe a bimodal pattern. Specifically, 383 bot accounts (52.4\,\%) were created less than one week before posting, contributing 24,954 tweets. On the other hand, 330 accounts (around 45\,\%) were over one year old, responsible for 197,680 tweets (around 88\,\% of all automated content). Almost no mid-age accounts were observed (only 18 accounts between 1 week and 1 year old), suggesting the prevalence of freshly created sockpuppets or compromised/purchased aged accounts. This modus operandi reflects established patterns in illicit ecosystems, where such services are readily available on cybercrime marketplaces \cite{lykousas2023cynicism}.

To better understand the differentiating attributes of these two types of accounts, we examine their number of followers. Expectedly, the newly created sockpuppets had practically no followers,  ($\bar{x}=0.1$),
 while the older accounts averaged 65 followers. Despite this disparity, the two groups achieved nearly the same aggregate reach: sockpuppets' tweets amassed 2,110,270 views (24,954 tweets), and older accounts' tweets 2,213,699 views (197,680 tweets), reaching a total of 4,323,969 views combined. Engagement, however, was effectively zero for both account categories (zero median retweets and favorites). This indicates that visibility is not driven by follower networks, but rather by demand-side discovery of a narrow set of coded CSAM keywords/hashtags. In practice, the campaign's spam reaches far more people than the bots' almost‑nonexistent follower bases would predict, with clear implications for recruiting and funneling new potential users into the ecosystem.

\begin{figure*}[!htb]
    \centering
    \includegraphics[width=0.7\textwidth]{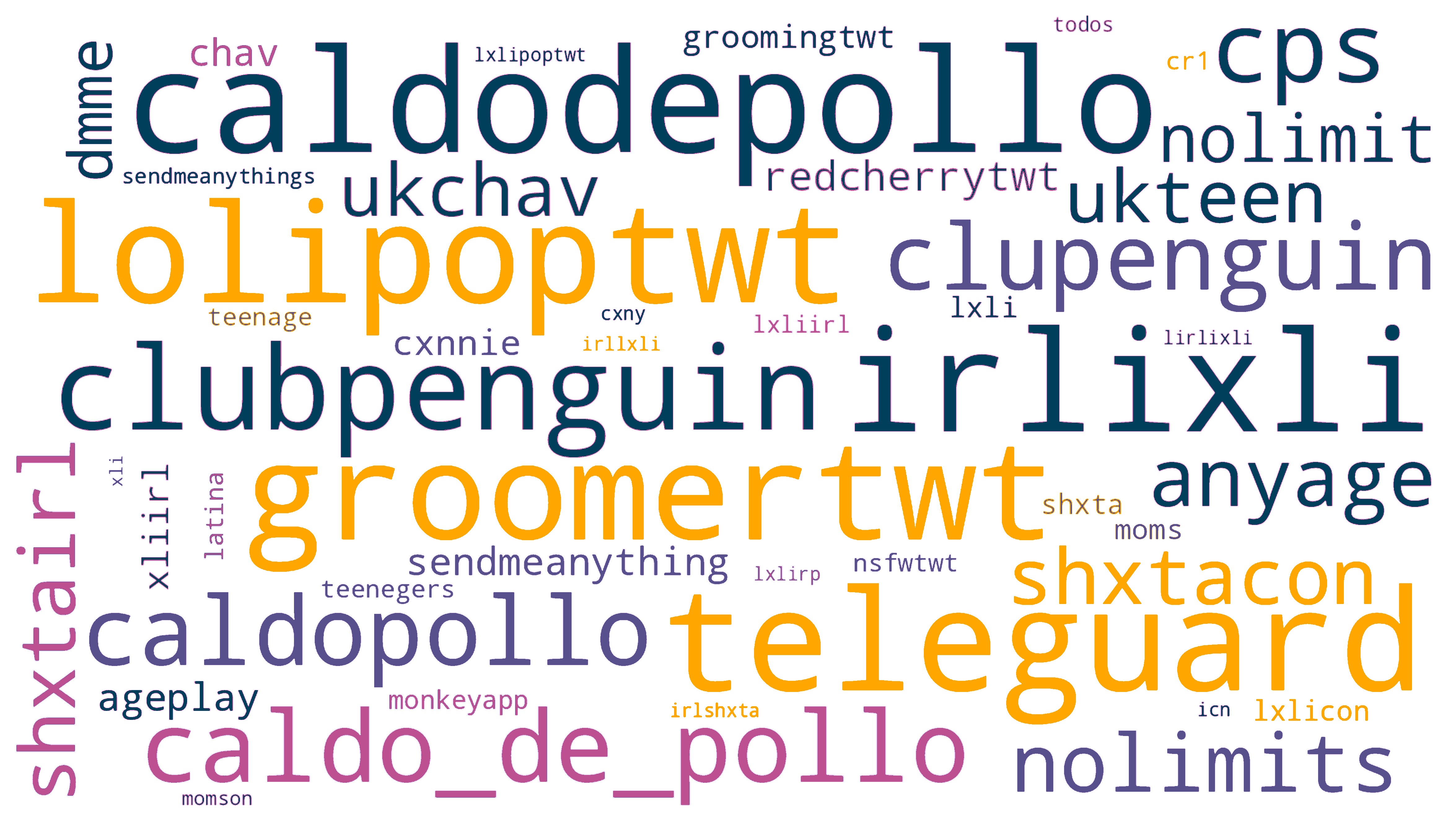}
    \caption{Hashtags used for disseminating the campaign on $\mathbb{X}$.}
    \label{fig:hashtags}
\end{figure*}

Next, we analyze the hashtags found in the collected tweets. In total, there are 298,958 hashtag occurrences, and in almost every tweet, multiple hashtags were used. We plot hashtags with more than five occurrences as a word cloud in Figure \ref{fig:hashtags}. We observe that beyond the most prominent tags which were explicitly advertised directly in the CSAM platforms, i.e., \texttt{\#caldodepollo} (45,624 occurrences), \texttt{\#irlixli} (38,311 occurrences) and \texttt{\#clubpenguin} (25,755 occurrences), there are hashtags with explicit terms like \texttt{\#lolipoptwt} (32,082 occurences), \texttt{\#groomertwt} (29,672 occurences) and \texttt{\#anyage} (4,555), as well as references to TeleGuardI\footnote{\url{https://teleguard.com}} (\texttt{\#teleguard} with 38,112 occurences), another possible channel of CSAM distribution. Finally, the existence of variations/misspellings like \texttt{\#caldo\_de\_pollo} (10,890 occurrences), \texttt{\#caldopollo} (9,422 occurrences), and \texttt{\#clupenguin} (8,609 occurrences) confirms the adoption of noisy text as a means to circumvent keyword-based detection mechanisms, a technique widely adapted by users involved in online grooming/CSAM ecosystems \cite{LykousasP21}.

Notably, 70.8\,\% of all tweets contained media attachments. This aligns with our earlier observations about AI-generated images containing embedded URLs and Telegram handles, used to bypass text-based moderation. Furthermore, 15\,\% of the tweets contained Telegram invite links, comprising 290 unique channels/accounts. The top channel was referenced in 4,665 tweets posted by eight different bot accounts.

\section{Analyzing the referral network}
\label{sec:network}
Delving into the collected JavaScript code from the CSAM platforms in our dataset, we were able to extract several important findings. For instance, most of the platforms deployed, even across different domains, used the same backend server, and, as a result, the same APIs.
Interestingly, consistent across most deployments is the use of a specific API that displays a real-time activity feed of the referral network, showing who invited whom and each user's accumulated invite count.  
This public ticker mechanism was probably designed to create engagement by motivating users to invite others to unlock access to more explicit CSAM, thus further spreading the campaign.

As the live expansion of the referral network was made available through this API, which did not require authentication, we queried it on a daily basis to collect information about inviters and invitees and to unveil the referral dynamics. This data collection effort also lasted for the same 9-month duration, but it should be noted that there were sparse periods of downtime related to the specific API, which in total account for 57 days where no data was collected, resulting in a complete snapshot of the referral network formation over 247 days.

\begin{figure*}[!htb]
    \centering
    \begin{subfigure}[b]{0.48\textwidth}
    \includegraphics[width=\textwidth]{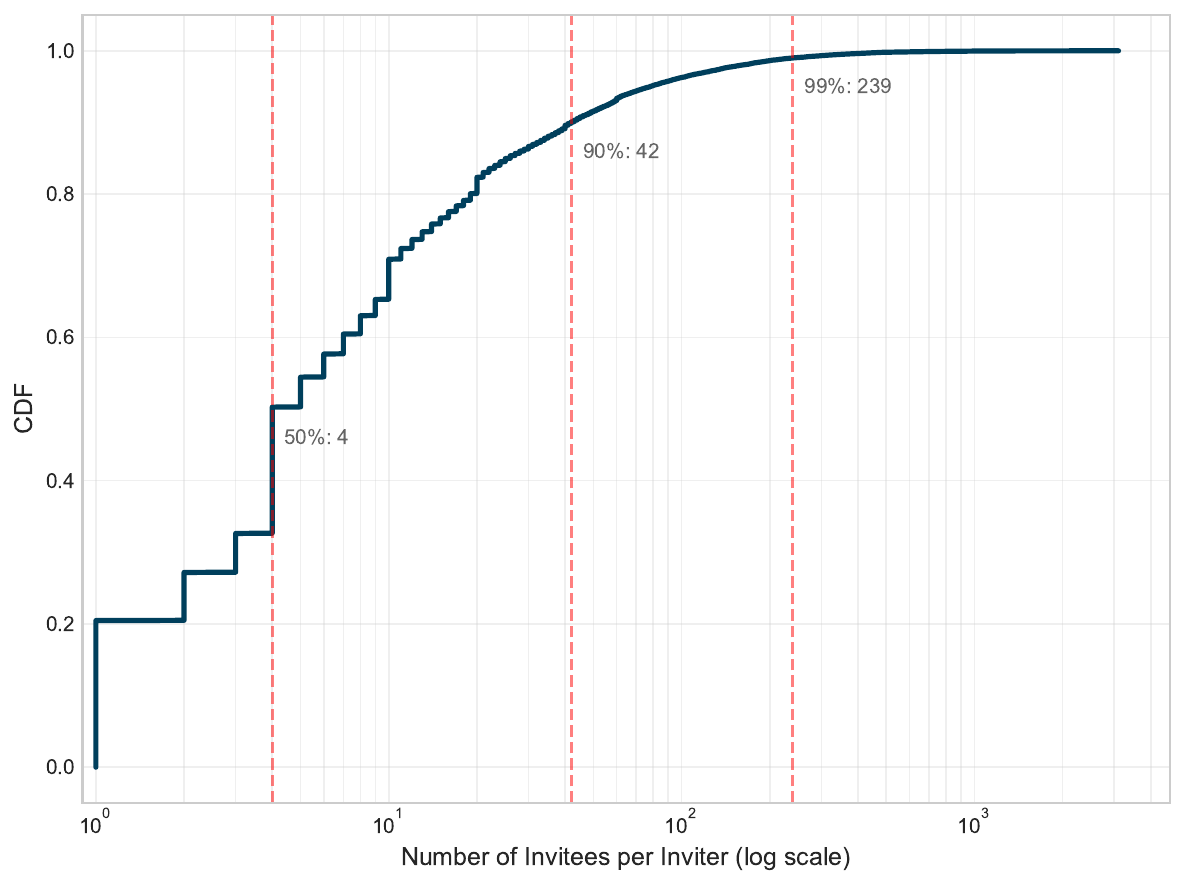}
    \caption{Cumulative distribution of invitees per inviter.}
    \end{subfigure}
    \begin{subfigure}[b]{0.48\textwidth}
    \includegraphics[width=\textwidth]{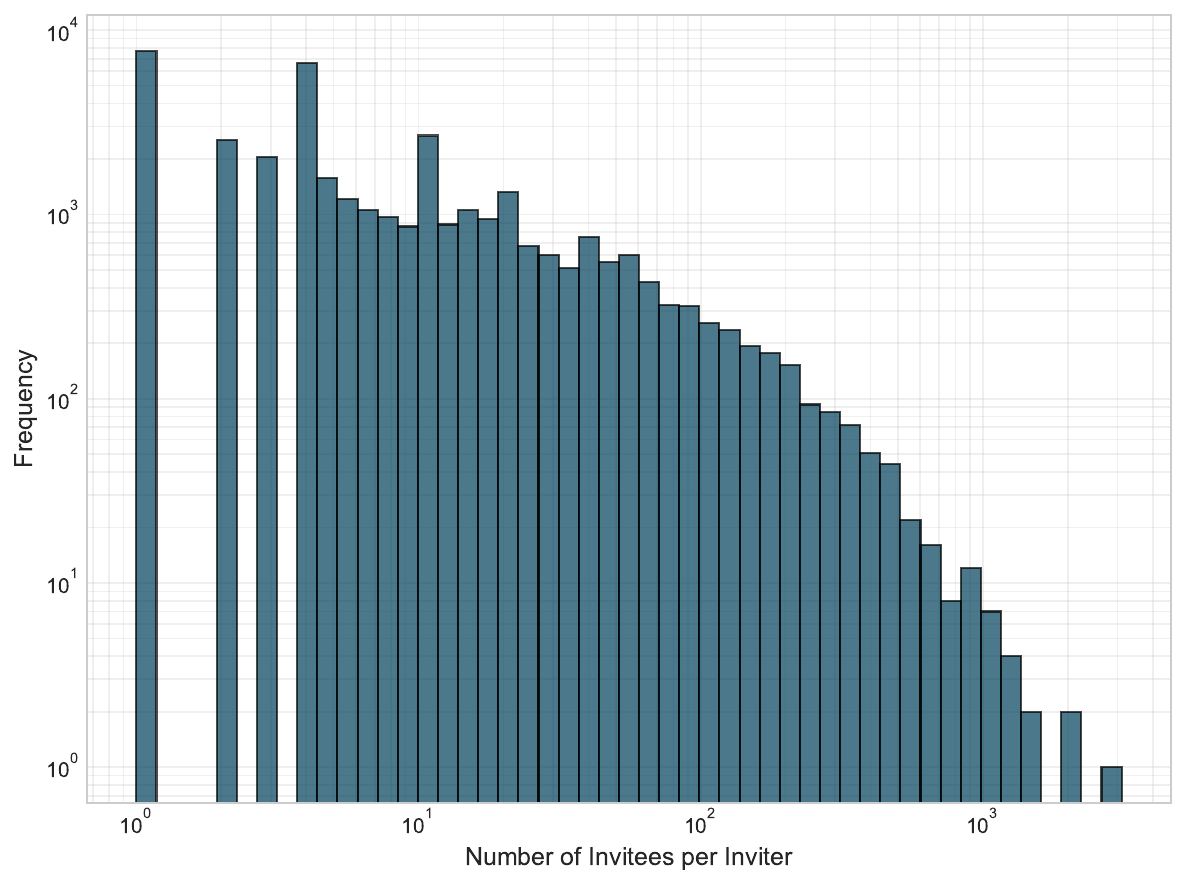}
    \caption{Distribution of invitees per inviter.}
    \end{subfigure}
    \caption{Invitees per inviter distribution.}
    \label{fig:inviter_dist}
\end{figure*}

In the collected data, we observed 700,531 new user registrations via the referral system, which were generated by 37,755 unique inviters. Notably, 22,431 invited users (3.2\,\%) became inviters themselves, further cascading the referral chain. We construct the inviter-invitee graph and plot the cumulative distribution function (CDF) and the frequency distribution of invitees per inviter on a logarithmic scale in Figure \ref{fig:inviter_dist}. We observe that 50\,\% of inviters recruited four or fewer users, 90\,\% recruited 42 or fewer, yet the top 1\,\% (377 users) recruited more than 239 users each. Specifically, the top 1\,\% of inviters generated 22.5\,\% of all registrations, while the top 10\,\% (3,775 users) accounted for 64.5\,\% of the campaign's network growth. Furthermore, the most prolific inviter single-handedly recruited 3,103 users. The frequency distribution is consistent with the power law distributions observed in scale-free networks \cite{barabasi1999emergence}, indicating that the observed recruitment/engagement of CSAM users is organic, meaning that the registered users are real, rather than artificial. Concretely, the distribution reveals three distinct user categories: (1) casual participants comprising the majority who recruit fewer than 10 users, likely seeking content access; (2) motivated affiliates in the top 10\,\% driving substantial growth; and (3) super-spreaders who may be campaign operators or highly incentivized distributors.

\section{Characterizing CSAM users}
\label{sec:consumers}
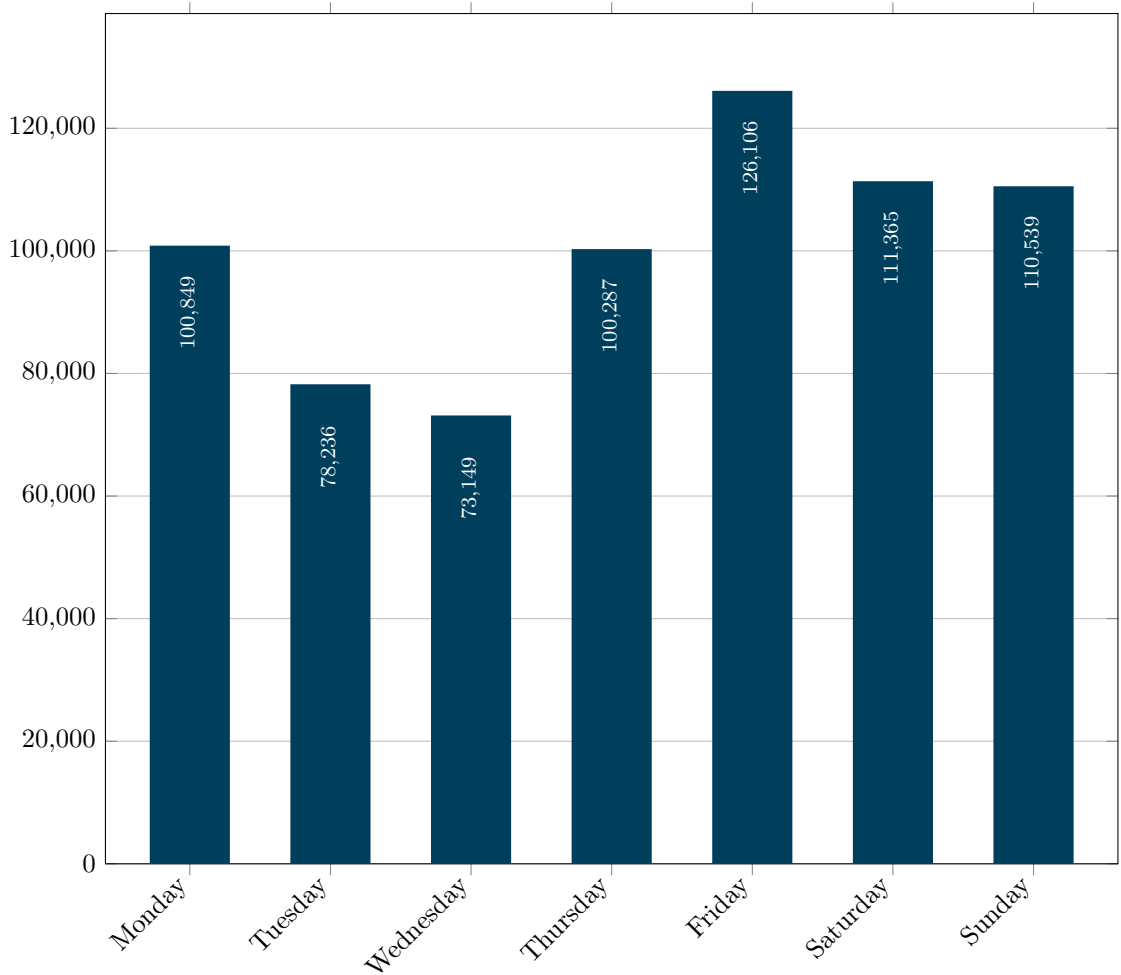
\begin{figure}[!ht]
    \centering
    \begin{tikzpicture}
\begin{axis}[
    ybar,
    ymajorgrids=true,
    bar width=30pt,
    width=\columnwidth,
    symbolic x coords={
        Monday, Tuesday, Wednesday, Thursday, Friday, Saturday, Sunday
    },
    xtick=data,
    xticklabel style={rotate=45, anchor=east},
    nodes near coords,
    ymin=0,
    scaled y ticks=false,
    yticklabel style={
        /pgf/number format/fixed,
        /pgf/number format/precision=5
    },
    every node near coord/.append style={
        rotate=90,
        xshift=-1.5cm,
        anchor=west,
        font=\footnotesize,
        /pgf/number format/fixed, 
        /pgf/number format/precision=0,
        color=white, 
    },
]

\addplot [fill=color1, draw=none] coordinates {
    (Monday,100849)
    (Tuesday,78236)
    (Wednesday,73149)
    (Thursday,100287)
    (Friday,126106)
    (Saturday,111365)
    (Sunday,110539)
};

\end{axis}
\end{tikzpicture}

    \caption{Total user registrations per day of the week.}
    \label{fig:registrations}
\end{figure}

Next, we analyze the data to extract behavioral patterns and characteristics of the users participating in the CSAM referral network. On average, $\approx$2,836 new users joined per day in our 247-day dataset, indicating sustained demand for this illegal content. In Figure~\ref{fig:registrations}, we plot the total number of registered users per day. Evidently, the most popular registration days are Friday, Saturday, and Sunday, with 348,010 total registrations. These three days average 116,003 registrations per day compared to 88,130 for other weekdays.

As we analyzed the data, we identified a brief window during which the API leaked additional user information, providing us with further insight into user characteristics. For approximately 30 days during our data collection period, the returned data regarding registration contained additional fields for each user, including their country, timezone, and User Agent. Although this appears to be a catastrophic operational security failure by the campaign operators, it provided invaluable data about the user base. In total, we captured these attributes for 65,261 users before the leak was patched.

\begin{figure*}[!th]
    \centering
    \includegraphics[width=\textwidth,trim={0 2cm 0 2cm},clip]{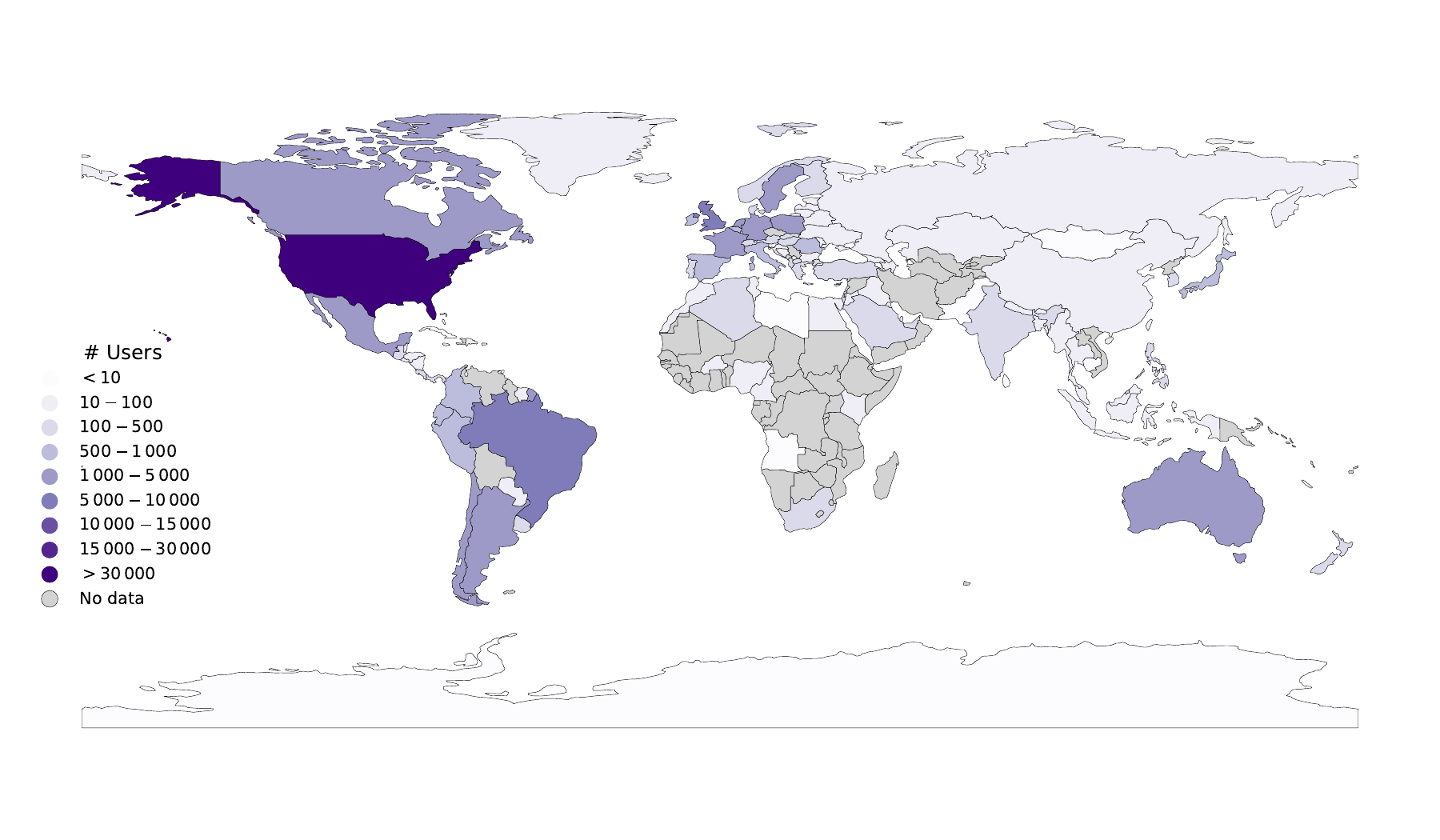}
    \caption{Geographic distribution of CSAM platform users.}
    \label{fig:visitors_map}
\end{figure*}

We present the geographic distribution of the users of the CSAM platform in Figure~\ref{fig:visitors_map}. The United States dominates with 38,382 users (58.8\,\% of the total with country attribute), followed by a second level comprising the United Kingdom (6,085), Brazil (6,079), Germany (3,513), Mexico (3,320), and Canada (3,046). In particular, Western Europe, the Americas, and Australia account for the vast majority of users, while regions with lower Internet adoption or restrictive policies show proportionally fewer participants. However, the users of the campaign platforms during the leak period were distributed in 110 different countries, indicating that this campaign has a global reach. A limitation of this analysis is that we cannot determine the extent of proxy or VPN/Tor usage, which may impact the actual geographic distribution of users.

\begin{figure}[!htb]
    \centering
    \begin{tikzpicture}
\begin{axis}[
    ybar,
    ymajorgrids=true,
    bar width=40pt,
    width=\columnwidth,
    symbolic x coords={
     Android, iOS, Linux,macOS,  Other, Windows
    },
    xtick=data,
    nodes near coords,
    ymin=0,
    scaled y ticks=false,
    yticklabel style={
        /pgf/number format/fixed,
        /pgf/number format/precision=5
    },
]

\addplot [fill=color1, draw=none] coordinates {
    (Android,22931)
    (iOS,31893)
    (Linux,988)
    (macOS,1610)
    (Other,1051)
    (Windows,6648)
};

\end{axis}
\end{tikzpicture}
    \caption{Operating Systems used by CSAM users.}
    \label{fig:os_dist}
\end{figure}
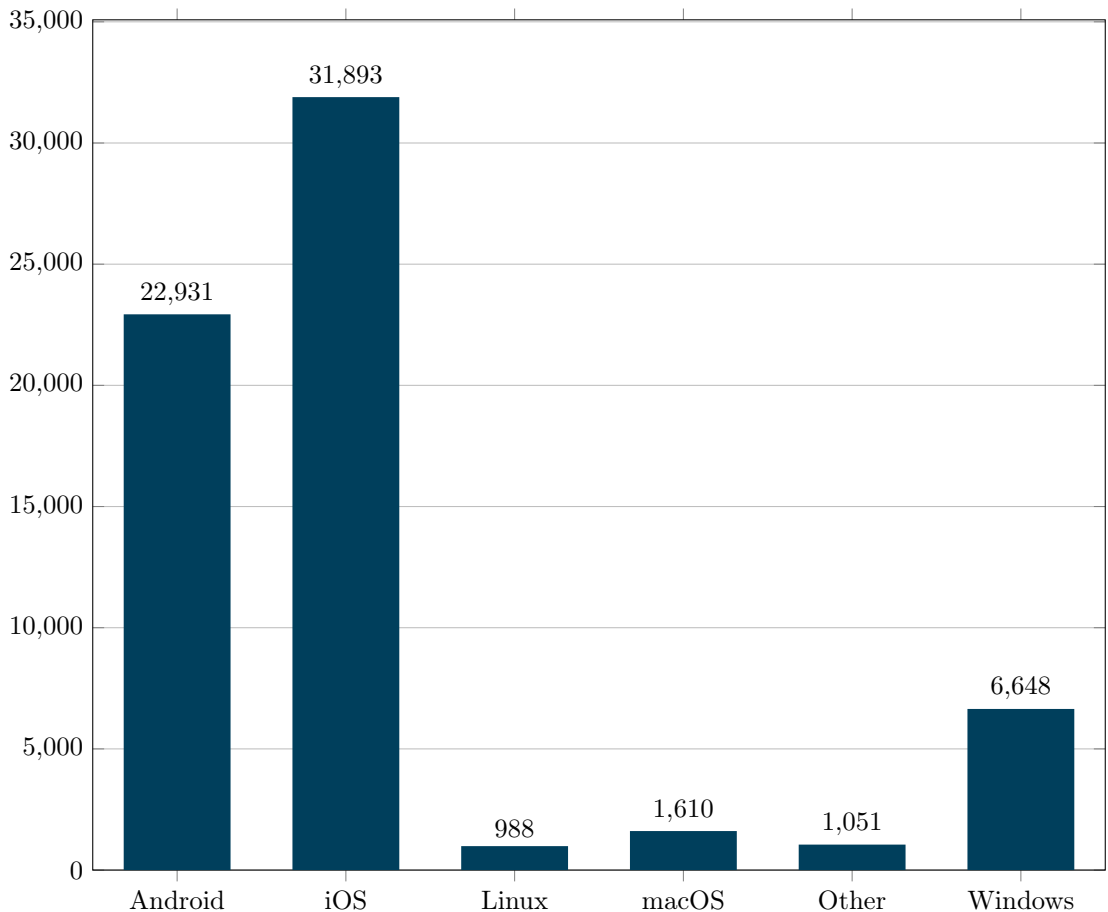

To understand the devices used to access these platforms, we parsed the User Agent strings using the \texttt{user-agents}\footnote{\url{https://pypi.org/project/user-agents/}} Python library. Fig.~\ref{fig:os_dist} shows the distribution across operating systems. We observe that almost half of the users were using iOS devices, followed by Android. Overall, mobile devices accounted for 84.1\,\% of usage. In contrast, desktop users account for a small fraction, as the combined share of desktop operating systems of Windows, macOS, and Linux represents less than 15\,\% of users. 
Finally, we tested the usernames in the complete dataset against a regular expression matching email patterns, identifying 1,606 distinct email addresses among the usernames. This means that, despite the criminal nature of the content, a subset of users still used email addresses for registration, even though the platform did not require or validate them. Thus, everyone could use arbitrary emails, even ones that did not belong to them.

\section{Limitations}
\label{sec:limitations}
Despite the scale and depth of our findings, this study is subject to several limitations, which, for transparency, we list below.

First, our data collection was constrained by the visibility of indexed content on public platforms and the limitations of API-based querying. While our methodology identified more than 1,000 domains and captured thousands of tweets and site metadata, it is highly likely that we only observed a fragment of the entire CSAM ecosystem operating on the clear web. Domains hosted on private forums, non-indexed sites, or behind paywalls may have been missed by our detection pipeline.

Moreover, several behavioral insights, such as user geography and browser data, were derived from a temporary API misconfiguration that leaked identifiable metadata. Although these data were collected passively and ethically, the leak period spanned only 30 days, representing less than 15\,\% of our total collection window. As such, the resulting characterization of users may not be applicable to the entire user base.

As discussed, to estimate the use of bots on social media platforms, we employed a conservative rule: accounts posting two or more tweets within a 60-second window were labeled as bots. While this minimizes false positives, it may underestimate the true scale of bot activity and overlook more sophisticated automation techniques that simulate human behavior. 

Finally, in the same vein of thought, the structure of the referral system suggests genuine viral spread; however, it is possible that some users fabricated referrals to unlock higher content tiers, thereby inflating the number of invites. Similarly, the use of arbitrary or fake email addresses during registration undermines any direct mapping between user identities and real individuals.

\section{Conclusion}
In this paper, we presented the first longitudinal, data-driven investigation of a large-scale CSAM referral campaign operating openly on the clear web. Our research shows that CSAM distribution is a persistent and evolving threat that plagues traditional social media platforms, whose content moderation and reporting mechanisms are easily circumvented. Moreover, the large-scale bot activity facilitates the global dissemination of content. The fact that the campaign has been operating for so long on the clear web rather than hidden on encrypted or anonymized platforms illustrates both the limitations of current enforcement mechanisms and the existence of a large, engaged user base. 

By combining multiple data collection pipelines and exploiting operational flaws in the infrastructure of CSAM campaigns, we were able to reveal in detail the structure and behavior of this ecosystem, from domain lifecycles and user acquisition funnels to device/browser usage patterns and geolocation data, without directly accessing illicit material. Our work provides evidence that a multi-level marketing-style system fuels much of the CSAM circulation on the clear web, and that the majority of engagement appears to stem from genuine users, rather than synthetic or inflated interactions. Nonetheless, our research reveals that illicit campaigns encounter operational issues that can be leveraged to gain deep insight into what is exchanged and to better understand their modus operandi, and aid future research and law enforcement efforts.

We argue that the huge and continuous increase in the CSAM distribution, coupled with the use of AI, will exacerbate the problem. The growing realism of synthetic content raises new questions regarding legality, ethics, and the identification of victims. As a result, law enforcement authorities will struggle to determine whether there is an actual victim in the seized content and identify it. 

These developments underscore the urgency for improved technical safeguards, rapid response pipelines, and coordinated action between platforms, researchers, and regulatory bodies. We hope that our work serves as both a warning and a foundation for future work. Continued interdisciplinary action is essential if we are to curb the proliferation of CSAM as new content is continuously generated and distributed, affecting thousands of victims worldwide.

\section*{Acknowledgment}
This work was partially supported by the European Commission under the Horizon Europe Programme, as part of the project SafeHorizon (Grant Agreement No. 101168562). 

The content of this article does not reflect the official opinion of the European Union. Responsibility for the information and views expressed therein lies entirely with the authors.

\end{document}